\documentstyle[preprint,aps]{revtex}

\def \nonu {\nonumber}

\begin{document}
\draft

\title {\bf THREE FLAVOUR QUARK MATTER IN CHIRAL COLOUR
DIELECTRIC MODEL }

\author{\bf Sanjay K. Ghosh and S. C. Phatak }
\address {Institute of Physics, Bhubaneswar-751005, INDIA.}
\maketitle

\begin{abstract}
We investigate the properties of quark matter at finite density
and temperature using the nonlinear chiral extension of Colour
Dielectric Model (CCM). Assuming that the square of the meson
fields devlop non- zero vacuum expectation value, the thermodynamic
potential for interacting three flavour matter has been
calculated. It is found that $<K^2>$ and $<\eta^2>$ remain zero
in the medium whereas $<\pi^2>$ changes in the medium. As a
result, $u$ and $d$ quark masses decrease monotonically as the
temperature and density of the quark matter is increased.In the
present model, the deconfinement density and temperature is
found to be lower compared to lattice results. We also study the
behaviour of pressure and energy density above critical
temperature.
\end{abstract}
\vfil \eject
\narrowtext

\section{ {\bf Introduction}}

The quark structure of hadrons suggested the possibility of a
transition from nuclear to quark phase at high density and/or
temperature. This indicates that there can be a phase
transition, usually termed as deconfinement transition,
inside neutron stars at very high densities so that neutron
stars may now consist of quark cores. On the other hand the
heavy ion collision experiments can also produce quark matter.
In fact the increasing availability of high energy beams is
giving new dimension to the quest for quark gluon plasma.
But the scenario gets complicated because of Chiral symmetry.
At low density or temperature chiral symmetry is broken and
eventually at some high density or temperature it is supposed to be
restored. So with increase in density and/or temperature there
can be two transitions, one is the deconfined transition where
nucleons dissolves into quarks and the other is chiral
transition where chiral symmetry is restored.

QCD is the theory of strong interaction. So one would like to
understand the properties of matter at extreme conditions and
the possible phase transitions using QCD. At present the only
method of such investigation is lattice calculation. The best
established results of lattice calculations [1] show that though
the energy density attains an ideal gas value very quickly above
deconfined transition temperature $T_c$, it takes a much larger
value of $T$, for the pressure, to attain the same behaviour. This
is an indication of nonperturbative behaviour which is poorly
understood. The character of the phase transition depends
sensitively on the number of light quark flavours. Although
there are large uncertainties, the current status is that for 2-
flavour light quarks, the transition is second order, while for
3 or more flavours it appears to be first order. But then
lattice calculation has its own complicacy, besides the problem
of large computer time needed. Till today these are limited to
systems with zero chemical potentials.It is unclear whether
strange quarks should be treated as light or heavy. Also one
should note that, in the real world since the mass of the light
quarks are not strictly zero, the second order phase transition
for light quark flavours may turn into a rapid cross over.

One can investigate the matter at extreme conditions, without
so much of complicacy, by using QCD motivated models, which have
been devloped and successfully applied to the study of the
properties of baryons. One can get a simple physical
understanding of the different characteristics of matter. There
is one more advantage of using such models.
In lattice, it is difficult to represent the chiral symmetry of
the continuum. On the other hand, in models one can
implement chiral symmetry exactly.
Many such investigations have already been done [2- 6].
Recently, there are fresh attempts to understand the chiral
symmetry transition using linear $\sigma$ model [7, 8]. In ref. [7]
authors have used $SU(3) \times SU(3)$ linear $\sigma$ model
Lagrangian in terms of the $3\times3$ matrix field $M$. In fact
this Lagrangian corresponds to the action with $M$ as order
parameter. They find that for realistic meson masses, the chiral
transition becomes a crossover phenomena, i.e the phase
transition behave smoothly in all thermodynamic quantities. With
the decrease in pion mass, the order of the transition changes
from crossover to first order and with
the decrease in $\sigma$ mass, it changes from crossover to
second order.
Also at the transition point only the $SU(2) \times SU(2)$ part of
the chiral symmetry is restored. Some authors have studied the
temperature dependence of chiral condensate from an interacting
pion gas, by using the fact that the condensate is related to
the derivative of the free energy density with respect to the
bare quark mass [9].  They find a critical temperature $T_c \sim
\sqrt{2}f_\pi$.
\par
In our earlier calculation [5], we have
used nonlinear chiral colour dielectric model (CCDM) to study
the properties of two flavour quark matter at finite temperature
and density. There we have shown that with our ansatz of nonzero
$F_{\pi}= <{\vec\pi}^2>/f_{\pi}^{2}$ and vanishing $<\vec\pi>$,
chiral symmetry is restored gradually
at very high temperature $T \sim (2- 3)T_c$, $T_c$
being the deconfinement temperature which is around 100- 120
MeV. This model has been applied further to study the properties
of stars [10, 11]. In the present paper we have extended our ansatz to
study three flavour matter at finite density and temperature.
Now our ansatz implies a nonzero $F_{\phi}=$
$<\phi^2>/f_{\pi}^{2}$, where $\phi$ corresponds to the meson
fields i.e. pion ($\pi$), kaon ($K$), or eta ($\eta$) [10- 12]. We find
that in the medium $F_K$ and $F_{\eta}$ are
zero where as $F_{\pi}$ attains a nonzero value for the range of
density and temperature we have considered. This implies that
strange quark mass remains constant in the medium where as $u$ and
$d$ quark masses decrease with density. So the chiral transition
occur only in the two flavour sector in our model. Also the
meson excitations play almost no role in determining the
properties of quark matter.

The paper is organised as follows. In section 2 we define the
CCDM and give a brief outline of our ansatz, most of the detail
being left for appendix. In section 3, we present the results of
our calculation and finally conclusion is given in section 4.

\section{\bf The Model }

The Colour Dielectric Model (CDM) is based on the idea of Nelson
and Patkos [13]. In this model one  generates the confinement of
quarks and gluons
dynamically through the interaction of these fields with scalar
field or colour dielectric field $\chi$. In the present work we
have used the chiral extension [10, 11, 14] (CCDM) of this model
to calculate quark matter EOS. The Lagrangian density of CCDM
is given by
\begin{eqnarray}
L(x)&= \bar\psi(x)\big \{ i\gamma^{\mu}\partial_{\mu}-
(m_{0}+m/\chi(x) U_{5}) + (1/2) g
\gamma_{\mu}\lambda_{a}A^{a}_{\mu}(x)\big \}\psi \nonumber \\
&+f^{2}_{\pi}/4 Tr ( \partial_{\mu}U
\partial^{\mu}U^{\dagger} ) -
1/2m^{2}_{\phi} \phi^{2}(x) -(1/4)
\chi^{4}(x)(F^{a}_{\mu\nu}(x))^{2} \nonu \\ &+ (1/2)
\sigma^{2}_{v}(\partial_{\mu}\chi(x))^{2}- U(\chi)
\end{eqnarray}
where $U = e^{i\lambda_{a}\phi^{a}/f_\pi}$ and $U_{5} =
e^{i\lambda_{a}\phi^{a}\gamma_{5}/f_{\pi}}$, $\psi(x)$,
$A_{\mu}(x)$, and $\chi(x)$ and $\phi(x)$ are quark, gluon,
scalar ( colour dielectric )and meson fields respectively.
$m$ and $m_{\phi}$
are the masses of quark and meson, $f_{\pi}$ is the pion
decay constant, $F_{\mu\nu}(x)$ is the usual colour
electromagnetic field tensor, g is the colour coupling constant
and $\lambda_{a}$ are the Gell-Mann matrices.  The flavour
symmetry breaking is incorporated in the Lagrangian through the
quark mass term $(m_{0}+m/\chi U_{5})$, where $m_0= 0$ for u and
d quarks. So  masses of u, d and s quarks are $m$, $m$ and
$m_{0}+ m$ respectively. So for a system with
broken flavour symmetry i.e. flavour asymmetric matter
, strange quark mass will be different from u and d quark masses.
The meson matrix is then consist of a singlet $\eta$, triplet of
$\pi$ and quadruplet of $K$. One can also describe a flavour
symmetric matter using above Lagrangian. For such systems, $m_0=
0$ and all the quark masses become same. The meson matrix $\Phi$
then becomes a eight component field. The self interaction
$U(\chi)$ of the scalar field is assumed to be of the form
\begin{eqnarray}
U(\chi)~=~\alpha B
\chi^2(x)[
1-2(1-2/\alpha)\chi(x)+(1-3/\alpha)\chi^2(x)]
\end{eqnarray}
so that $U(\chi)$ has an absolute minimum at $\chi = 0$ and a secondary
minimum at $\chi = 1$. The interaction of the scalar field with quark
and gluon fields is such that quarks and gluons can not exist in the
region where $\chi= 0$.
In  the  limit  of  vanishing  meson  mass,  the
Lagrangian of eqn.(1) is invariant under chiral transformations of
quark and meson fields.
\par
There are two different approaches used when chiral models are
applied for the study of hadrons. One is cloudy bag model [15],
which is a perturbative method. The another one is hedgehog
approach [16], which is nonperturbative but is not applicable for
infinite matter. In our calculation, in an attempt to go beyond
the perturbative approach of cloudy bag model, we assume that
because of nonvanishing quark and antiquark densities, the squre
of the expectation value of meson fields devlop a nonzero
value i.e. $<\phi^{2}>\not = 0$. On the other hand we assume
that expectation value of the meson field vanishes in the medium
. For an infinite system of quarks we can assume that
$<\phi^{2}>$ is independent of space and time. The meson
excitations are then defined in terms of the fluctuations about
$<\phi^{2}>$, so that $\phi^{2}~=~<\phi^{2}>~+ ~{\phi^{'}}^{2}$.
Defining $F_{\phi}~= <\phi^{2}>/f_{\pi}^{2}$, the CDM Lagrangian
can be rewritten in terms of $F_{\phi}$s and meson excitations
$\phi^{'}$. In our calculation the scalar field $\chi$ and
$F_{\phi}$ have been calculated in the mean field approximation
and quark- gluon, gluon- gluon and quark- meson excitations are
treated perturbatively. Thus, keeping the terms upto two meson
exchange interaction we get,
\begin{eqnarray}
L(x)&=\bar\psi(x)\big \{ i\gamma^{\mu}\partial_{\mu}-
(m_{0}+m/\chi(x) f_{1}(F_{\phi})) + (1/2) g
\gamma_{\mu}\lambda_{a}A^{a}_{\mu}(x)\big \}\psi \nonumber \\
&-(1/4)\chi^{4}(x)(F^{a}_{\mu\nu}(x))^{2}- U(\chi)+ (1/2)
\sigma^{2}_{v}(\partial_{\mu}\chi(x))^{2}
-{1 \over {2}}
m_{\phi}^{2}f_{\pi}^{2}F_{\phi}) \nonu \\
&+{1\over {2}}f_{2}(F_{\phi})(\partial_{\mu}(x))^2 -1/2m^{2}_{\phi}
{\phi^{'}}^{2}(x)
-{{im}\over {f_{\pi}{\chi}}}\bar\psi(x)\gamma_{5}f_{I1}(F_\phi)
\psi(x) \nonu \\ &+ {m\over {2f_{\pi}\chi}}\bar\psi(x)f_{I2} \psi(x)
\end{eqnarray}
The last two terms are the interactions of meson excitations with quarks.
The $f_1$, and $f_2$, are the $3\times
3$ diagonal matrices , each term of which is a polynomial in $F_\phi$s'
i.e. $F_\pi$, $F_K$, $F_\eta$ and their cross terms. $f_{I1}$
and $f_{I2}$ are interaction matrices for one meson
and two meson exchanges respectively. These are also
$3\times 3$ matrices but are not diagonal. In appendix
, we have given the method of evaluating the $f_{1}$, $f_{2}$,
$f_{I1}$ and $f_{I2}$ and also the expressions of their nonzero
contributions for three flavour asymmetric
quark matter, symmetric matter and two flavour quark matter.
For three flavour (both flavour symmetric and
asymmetric ) the above polynomials
 are not summable, but are fast convergent and terms upto
3rd order gives reasonably good convergence. On the other hand,
these polynomials are simpler for two flavour matter and can be summed
to a compact form as given in appendix and ref. [5]. In our
calculations scalar field $\chi$ and $F_\phi$ have been
evaluated in the mean field approximation and quark- gluon and
quark- meson interactions are treated perturbatively. In this
approximation, $\chi$ is independent of space
and time for uniform quark matter. In actual calculation,
the mean field values of
$\chi$ and $F_\phi$ are determined by minimizing the thermodynamical
potential $\Omega$.
Here, to have a qualitative feeling of
the different $F\phi$s', let us consider the equation of motion
for $F_\phi$ at zero temperature and without interactions.
\begin{eqnarray}
(i\gamma^{\mu} \partial_{\mu}~-~(m_0+{m\over {\chi}}f_1(F_\phi))\psi(x)
{}~=~0
\end{eqnarray}
\begin{eqnarray}
U'(\chi)~=~{m\over{\chi^2}}f_{1}(F_\phi) {\gamma\over
{2\pi^2}}\int^{k_f}_{0} k^2 dk m^{*}/{(k^2+{m^{*}}^{2})}^{1/2}
\end{eqnarray}
\begin{eqnarray}
f_2(F_\phi)\partial^{\mu} \partial_{\mu} \phi^{'}(x)~+~ m_{\phi}^{2}
\phi^{'}(x)~=~0.
\end{eqnarray}
\begin{eqnarray}
m_{\phi}^2f_{\pi}^2~=~{m\over{\chi}}f^{'}_{1}(F_{\phi}){\gamma\over
{2\pi^2}}\int^{k_f}_{0} k^2 dk m^{*}/{(k^2+{m^{*}}^{2})}^{1/2}
\end{eqnarray}
where, $m^{*}=
(m/\chi)(f_{1}(F_{\phi}))_{11 or 22}$ for $u$, $d$ quarks and
$m^{*}= m_0+(m/\chi)(f_{1}(F_{\phi}))_{33}$ for s quarks.
$k_f$ is the fermi momentum
and $\gamma$ is the spin- colour degeneracy factor for quarks.
The equation of motion for quarks (eqn.(4))shows that $m^*$ can
be identified
with the effective quark masses in the presence of $\chi$ and $F$.
Also the equation of meson fluctuations show that, for non zero
$F$ the energy momentum relation for meson fluctuation is
$E_{\phi}(k)= \sqrt{k^2+{m^{*}}^{2}_{\phi}}$, where $m^{*}_{\phi}=
{m_{\phi}\over {\sqrt {{f_{2}(F_{\phi})}}}}$. Thus the effective
pion mass
increases with $F$ and hence density in the medium. The mean field
values of $\chi$ and $F$ can be evaluated by solving the equations
(4) - (7) self consistently.

The characteristics of $F_{\phi}$ can be understood from eqn.(7). The
physical values of $F_{\phi}$ should be greater than zero. The
eqn.(7) yields a positive value of $F_\phi$ only if
${\gamma \over {2\pi^2}}\int^{k_f}_{0} k^2 dk m^{*}/{(k^2+{m^{*}}^{2})}^{1/2}
\ge {(m_{\phi}f_{\pi}\chi/m)}^2$, that is if $k_f$ ( or the quark
density) is larger than a certain value. Below this density, $F_{\phi}$
remains zero and, therefore, the eqn. (7) becomes redundant.
 Now since $m_\pi < m_K < m_\eta$
, the respective fermi momentum $k_f$ (for $\pi$) $<$ $k_f$(for $K$)
$<$ $k_f$(for $\eta$), i.e. as quark density increases, the $F_\pi$
will appear first. The $F_K$ and $F\eta$
will appear much later as the masses of $K$ and $\eta$ are
much higher than the pion mass. Also, $F_{\phi}$ increases with
quark density. This means that effective quark masses will decrease
with density and may vanish for a certain value of $F_{\phi}$ .

Using the modified Lagrangian we calculate thermodynamic potential
$\Omega$ upto second order in quark- gluon coupling and quark-
meson couplings [4, 5]. The thermodynamic
quantities are obtained from $\Omega$ using standard methods [4,
5]. The
mean field values of $\chi$ and $F_{\phi}$ are obtained by
minimizing $\Omega$ with respect to $\chi$ and $F_{\phi}$;
\begin{eqnarray}
{(\partial{\Omega}/\partial{\chi})}_{T,\mu}=~0~;~~~~
{(\partial{\Omega}/\partial{F_{\phi}})}_{T,\mu}= 0
\end{eqnarray}

\section{\bf Results and Discussion}

The thermodynamic properties of the quark matter are calculated
for a number of parameter sets of chiral CDM ($B$, $m$,
$\alpha$x and $g_s$) which reproduce baryon masses. Earlier
calculations show that these parameters are not determined
uniquely by the fitting procedure [14]. In particular it has been
found that good fits to baryon masses are obtained for
$0.6GeV\leq {m_{GB}}\leq 3GeV$, $m_{q}(u,~d)\leq 125MeV$,
$m_{q}(s)\sim 300MeV$ and $B^{1/4}\leq 150MeV$, where $m_{GB}=
\sqrt {2B\alpha/\sigma_{v}^{2}}$. On the other hand, if one
insist for good fits of charge radius and magnetic moments as
well, then the fits are better for lower values of $m_{GB}$,
$m_{q}(u, d)$ and $B^{1/4}$.
But, lower
bag pressure $B^{1/4}$ gives a lower value for transition density and
temperature. In fact in ref. [11] we found that for such parameter
sets one even gets a stable quark matter phase at nuclear matter
densities. So we have used parameter sets for which one gets
reasonable behaviour and  we find
that the results for these parameter sets are qualitatively
similar. Therfore we have considered a representative
parameter set ( $B^{1/4}= 152MeV$, $m_{u,d}= 92MeV$, $m_s=
295MeV$, $\alpha= 36$ and $g= 1.002$ ) for a detailed
discussion. Since the analytic derivatives of the thermodynamic
potential with respect to $\chi$ and $F_{\phi}$s are nor easy to
obtain, we have numerically minimized $\Omega(T, \mu)$ as
function of $\chi$ and $F_{\phi}$. We find that the mean field
values of $\chi$ remain close to $1$ for both two and three
flavour matter, where as, for three flavour quark
matter
$F_{K}$ and $F_{\eta}$ remain zero in the medium for the
range of density as well as temperature considered in our
calculation. $F_{\pi}$ is
zero for small values of $T$ and $\mu$ to about 2 or more for $T=
400MeV$ and $\mu= 500MeV$. From the expressions given in
appendix it can be found
that $u$ and $d$ quark couple with all the $F_{\phi}$s' i.e.
$F_{\pi}$, $F_{K}$ and $F_{\eta}$. On the other hand, $s$ quark
couples with only $K$ and $\eta$. So $F_{K}~=~F_{\eta}~=~0$
implies that strange sector of the $SU(3)\times SU(3)$ decouples
and strange quark mass remains constant in the
medium. $m_u$ and $m_d$ become small at large values of baryon
density $n_B$ and temperature $T$. So, in our model, chiral
symmetry restoration occurs for $SU(2)\times SU(2)$ sector only.
The variation of $F_{\pi}$ with $n_B$ and T are shown in Fig.1
(a and b) and Fig.2(a and b) respectively. Corresponding curves
for flavour symmetric matter are shown by the curves c and d. We
find that variation of quark mass with $n_B$ and $T$ is slower
for flavour symmetric matter.

The equation of motion for $F_{\pi}$ [eqn.(7)] shows that for
lower values of $m_{\pi}$, the density at which $F_{\pi}$ is
nonzero comes down. It is also obvious that one can not lower
$m_\pi$ arbitrarily as below a certain value of $m_{\pi}$, there is
no solution. In Fig.3, we have plotted quark mass
with $n_B$ for different $m_\pi$. We find that for lower
$m_\pi$, quark mass approaches zero faster i.e. the chiral
restoration occur at lower density for lower pion mass. Similar
behaviour is obtained for finite T and zero $\mu$ as well.
But the nature of
the chiral transition, in our calculation, does not change with
$m_\pi$ and chiral restoration phase is reached smoothly. This
behaviour is different compared to ref. [7], where authors find
that nature of transition changes from crossover to
first order with decreasing pion mass. In our model,
the effective pion mass increases with increase in density and
temperature. The temperature variation of $m_\pi$ is shown in
Fig.4 It shows that initially pion mass increases to about $150
MeV$ and then remain constant. As discussed earlier, as the
chiral behaviour of the $SU(3)\times SU(3)$ is described by
$SU(2) \times SU(2)$ part, the similar behaviour is observed for
two flavour matter as well [5].

Figure 5 shows the variation of pressure with density for
different values of temperature. At zero temperature the
pressure is negative for $n_B < 0.35fm^{_3}$. This density
defines the boundary of the quark phase [5]. The density at which
the pressure vanishes decreases with increasing temperature and
above a certain critical temperature ($\sim 110- 120MeV$ in our
model) the pressure is always positive. This temperature
can be defined as the critical temperature for the QGP phase
transition in our model. We find that critical temperature for
two flavour matter ($\sim 115MeV$) is higher than the three
flavour matter($\sim 110MeV$). For flavour symmetric case
the transition temperature is lowest($\sim 105MeV$). But, as
evident from the values quoted, the $T_c$ is not much sensitive
to the different composition of the quark matter system.
The best evidence from the lattice QCD calculations show that
for two dynamical flavours transition temperature is around
$150MeV$ and the transition density is around 4- 8 times the
nuclear matter density. For three flavour matter there are no
convincing calculations, but one can assume it to be not much less
than $150MeV$, keeping in mind the large value of $m_s$ compared
to $u$ and $d$ quarks. So, in our model, critical temperature
for deconfined transition is lower than the lattice QCD results.

For an ideal gas of quarks and gluons, $E/T^4= \pi^2/30(\gamma_g
+ 7/4\gamma_q = 15.63$ for three flavour matter and $P={1\over
3}E$. Lattice calculations [1] show that energy density increases
above the $T_c$ and quickly attains the ideal gas value. On the
other hand pressure needs a larger temperature($T\sim T_c$) to
reach the ideal gas value. This is one indication that
nonperturbative effects are still present, in the QGP, atleast
just above the critical temperature. The values of $E/T^4$ and
$3P/T^4$ for $\mu= 0$, from our model calculation, are given in
Fig.6. The figure shows that at high temperature, $E/T^4$
approaches a value close to $15$ and as the temperature is
lowered, there is a small increase in $E/T^4$. We do not show
the results below the $T_c$ as the calculation is not applicable
at these temperatures. Fig.6 also shows that $P/T^4$ is a
monotonically increasing function of temperature. We have also
fitted $E/T^4$ and $P/T^4$ with functions of the form $f(T)=
\sum {a_i/T^i}$ with $i$ runing from 0 to 4. We find that a good
fit to $E/T^4$ is obtained with $a_0= 14.38$, $a_1= 644.15$,
$a_2= -1.15\times 10^5$, $a_3= 0$ and $a_4= 9.22\times 10^8$.
For $P/T^4$, a good fit is obtained with $a_0= 5.68$, $a_1=
-292.12$, $a_2= 27753.10$, $a_3= 0$ and $a_4= -7.699\times
10^8$. The corresponding fits for two flavour matter is similar
to the one obtained in ref. [5]. Because of the constant background
of the scalar field $\chi$ and $F$, we expect that the energy
density and pressure should have a temperature independent term.

\section{\bf Conclusion}

The non-linear version of the chiral colour dielectric model is
used to study the properties of quark matter in the present
work. We have assumed that due to non- zero quark and antiquark
densities, the square of the meson fields ( pion, kaon and eta )
devlop a vaccum expectation value. In the matter, the $<K^{2}>$
and $<\eta^{2}>$ remain zero throughout the range of densities
and temperatures considered. On the other hand, $<\pi^{2}>$ has
a non-zero value in the medium. As a result, the $s$ quark
mass remain constant in the medium, where as, effective $u$ and
$d$ quark masses decrease continuously as the density and
temperature of the quark matter is increased. If we interpret
this as a restoration of chiral symmetry, our calculation implies
that chiral symmetry is restored smoothly in chiral colour
dielectric model. Also, the chiral restoration occurs in
$SU(2)\times SU(2)$ part only.  There is a small increase in
the pion mass with temperature and density in our model.
Furthermore, if the pion mass is decreased, the quark mass drops
faster i.e. the chiral restoration occurs for lower density and
temperatures. Here one thing should be noted that quark- meson
interactions play almost no role in determining the quark matter
characteristics as their contribution to $\Omega$ is much less
compared to pion background field and quark- gluon interaction
contribution [5].

The pressure of the quark matter  becomes negative at lower
densities and temperatures, which implies that a confined phase
is reached when the temperature or density of the quark matter
is lowered. We find that the critical temperature does not vary
much for different composition of the quark matter.
 The values of the critical temperature and density
at which deconfinement transition take place are smaller than
the values obtained in lattice calculations [1]. The behaviour of
the pressure and density calculated for $\mu~=~0$ is
quantitatively similar to that obtained in lattice calculations.
\vfill
\eject
\newpage

\appendix
\section{}

Here we will describe the method of calculating the Lagrangian
(3) in terms of the ${{<\phi^{2}>}\over {f_\pi}^{2}}~=~F_{\phi}$ and
the meson excitations $\phi^{'}$. The meson octet $\Phi$ is,
\begin{eqnarray}
\lambda_{a}\phi^{a}= \Phi= \left( \begin{array}{ccc}
{\pi^{0}\over \sqrt{2}}+{\eta\over \sqrt{6}} & \pi^{+} & K^{+}
\\
\pi^{-} & {-\pi_{0}\over \sqrt{2}}+{\eta \over \sqrt{6}} &
K^{0} \\
K^{-} & \bar K^{0} & {-2\eta \over \sqrt{6}}
\end{array} \right)
\end{eqnarray}
where $\lambda_{a}$ are the Gellmann matrices. The chiral
symmetry operator and meson kinetic energy operator are $U_{5}=
e^{i\gamma_{5}\Phi/f_{\pi}}$ and $U= e^{i\Phi/f_{\pi}}$.
General procedure is to expand the exponential and contract the
different fields following the different rules as prescribed
below.

\begin{eqnarray}
\pi^{+}~\pi^{-}= \pi^{-}\pi^{+}= {\pi^{0}}^{2}= <\pi^{2}>/j_{\pi}=
F_{\pi} f_{\pi}^{2}/j_{\pi}    \\
K^{+}~K^{-}= K^{-}K^{+}= K^{0}\bar {K^{0}}= \bar {K^{0}}K^{0}=
<K^{2}>/j_{K}= F_{K} f_{\pi}^{2}/j_{K}    \\
 \eta^{2}= <\eta^{2}>/j_{\eta}=F_{\eta} f_{\pi}^{2}/j_{\eta}
\end{eqnarray}
where in a flavour asymmetric matter the factor due to isospin
recoupling are $j_{\pi}= 3$ for pions, $j_{K}= 4$
for kaons and $j_{\eta}= 1$ for eta. For flavour symmetric
matter $j_{\pi}= j_{K}= j_{\eta}= 8$.

In general for a operator $O(\phi(x))$, for any $+$ve integer
$n$, we can write,
\begin{eqnarray}
O(\phi(x))~=~<O(\phi(x))>~ +~ \sum_{n}[O(\phi(x))]_{n}
\end{eqnarray}
where $[O(\phi(x))]_{n}$ implies that there are $n$ meson fields
($\phi^{'}(x)$) in the expression. For our present calculation,
we need terms upto $n= 2$. Let us now evaluate the expansion for
different combinations of meson fields.

\begin{eqnarray}
<(\pi^{+}\pi^{-})^{n}>~=~n<\pi^{+}\pi^{-}><(\pi^{+}\pi^{-})^{n-1}>\nonu
\\
=n(n-1)<\pi^{+}\pi^{-}>^{2}<(\pi^{+}\pi^{-})^{n-2}
\nonu \\
=~{n! \over {j_{\pi}^{n}}}<\pi^{2}>^{n}
\end{eqnarray}
where the factor $n$ implies that there are $n$ ways to contract
one $(\pi^{+}\pi^{-})$ out of $n$ $(\pi^{+}\pi^{-})$ fields.
Once first contraction is done, then second one has $n-1$
possibilities. Hence we get the factor $n(n-1)$. Similarly one
can get the form for other fields.
\begin{eqnarray}
<(\pi^{0})^{2n}>~=~(2n-1)<{\pi^{0}}^{2}><(\pi^{0})^{2n-2}> \nonu
\\ =~(2n-1)(2n-3)<\pi^{0}>^{2}<(\pi^{0})^{2n-4}> \\
=~{(2n-1)!! \over {j_{\pi}^{n}}}<\pi^{2}>^{n} \\
<(\eta)^{2n}>~=~{(2n-1)!!{j_{\eta}^{n}}}<\eta>^{n}  \\
<(K^{+}K^{-})^{n}>~=~{n!\over {j_{K}^{n}}}<K^2>^{n} \\
<(K^{0}\bar {K^{0}})^{n}>~=~{n! \over {j_{K}^{n}}}<K^2>^{n}
\end{eqnarray}

The two meson interaction term can be evaluated following
the above procedure.
\begin{eqnarray}
[(\pi^{+}\pi^{-})^{n}]_{2}~=~{n^{2}(n-1)! \over
{j_{\pi}^{n-1}}}<\pi^{2}>^{n-1} {\pi^{+}\pi^{-}}
\end{eqnarray}
The $n^{2}$ factor shows that there are $n\times n$ ways to
collect a $\pi^{+}\pi^{-}$ pair from $n$ pairs. The
remaining $n-1$ pairs can be contracted in $(n-1)!$ ways. For
other meson fields we get,
\begin{equation}
[{(\pi^{0})}^{2n}]_{2}~=~{n(2n-1) (2n-3)!! \over
{j_{\pi}^{n-1}}}<\pi^{2}>^{n-1} {{\pi^{0}}^{2}}
\end{equation}
\begin{eqnarray}
[\eta^{2n}]_{2}~=~{n(2n-1) (2n-3)!! \over
{j_{\eta}^{n-1}}} <\eta^{2}>^{n-1} {\eta^{2}}  \\
{[(K^{+}K^{-})^{n}]}_{2}~={n^{2}(n-1)! \over
{j_{K}^{n-1}}}<K^{2}>^{n-1} {K^{+}K^{-}} \\
{[(K^{0}\bar {K^{0}})^{n}]}_{2}~=~{n^{2}(n-1)! \over
{j_{K}^{n-1}}}<K^{2}>^{n-1} {K^{0}\bar {K^{0}}}
\end{eqnarray}
For odd $m$, $[\phi^{2n}]_{m}$ vanishes, since $<\phi>= 0$. Also
for odd powers, $<\phi \phi^2>= 0$. The one meson exchange terms
are calculated from the terms containing odd powers of the meson
concerned. For any $+$ve integer $n$ we can write,
\begin{eqnarray}
{[(\pi^{+})^{n}(\pi^{-})^{n+1}]}_{1}~=~(n+1)\pi^{-}<\pi^{+}\pi^{-}>^{n}
\nonu \\
{}~=~{(n+1) n! \over {j_{\pi}^{n}}}<\pi^2>^n\pi^{-}
\end{eqnarray}
where one $\pi^{-}$ can be picked up from the $n+1$ fields in
$n+1$ ways and rest of the $n$ pairs of $\pi^{+}\pi^{-}$ can be
contracted in $n!$ ways. Similarly,
\begin{eqnarray}
{[(\pi^{+})^{n+1}(\pi^{-})^{n}]}_{1}~=~{(n+1) n! \over
{j_{\pi}^{n}}}<\pi^2>^n\pi^{+}  \\
{[(\pi^{0})^{2n-1}]}_{1}~=~{(2n-1)(2n-3)!! \over
{j_{\pi}^{n-1}}}<\pi^{2}>^{n-1}\pi^{0}  \\
{[\eta^{2n-1}]}_{1}~=~{(2n-1)(2n-3)!! \over
{j_{\eta}^{n-1}}}<\eta^{2}>^{n-1}\eta  \\
{[(K^{+})^{n}(K^{-})^{n+1}]}_{1}~=
{}~=~{(n+1) n! \over {j_{K}^{n}}}<K^2>^{n}K^{-} \\
{[(K^{+})^{n+1}(K^{-})^{n}]}_{1}~=
{}~=~{(n+1) n! \over {j_{K}^{n}}}<K^2>^{n}K^{+} \\
{[(K^{0})^{n}(\bar {K^{0}})^{n+1}]}_{1}~=
{}~=~{(n+1) n! \over {j_{K}^{n}}}<K^2>^{n}\bar {K^{0}} \\
{[(K^{0})^{n+1} (\bar {K^{0}})^{n}]}_{1}~=
{}~=~{(n+1) n! \over {j_{K}^{n}}}<K^2>^{n} {K^{0}} \\
\end{eqnarray}

\bf {Case a: Three flavour asymmetric matter}

So starting from $U_{5}$, considering only the even powers we
get a $3\times
3$ diagonal matrix $(f_{1}(F_{\phi}))_{ii}= <U_{5}>$, with fully
contracted terms, which contributes to the quark mass terms.
\begin{eqnarray}
(f_{1}(F_{\phi}))_{1~1}~=~(F_{1}(F_{\phi}))_{2~2}~=~
1- {F_{\pi}\over 2!} + {5F^{2}_{\pi}\over {3~4!}} -
{35F^{3}_{\pi}\over {9~6!}}...........\nonu  \\
- {F_{k}\over {2~2!}} + {3F_{k}^2\over {4~4!}} - {3F_{k}^3\over
{2~6!}}.............\nonu \\
- {F_{\eta}\over {3~2!}} + {F_{\eta}^{2}\over {3~4!}} -
{5F_{\eta}^{3}\over {9~6!}}...........\nonu \\
+ F_{\eta}({2F_{\pi}\over {4!}} - {25F_{\pi}^{2}\over {3~6!}} +
.........) \nonu \\
-5~F_{\eta}^{2}{F_{\pi}\over {5!}}....
+ F_{k}({3F_{\pi}\over {2~4!}} - {25F_{\pi}^{2}\over {6~6!}} +
........ \nonu \\
-7~F_{k}^{2}{F_{\pi}\over {2~6!}}...
+F_{\eta}({F_{k}\over {2~4!}}- {3~F_{K}^{2}\over {2~6!}}...)
-3~F_{\eta}^{2}{F_{k}\over {2~6!}}...
+ O(F_{\phi}^{4}) \\
(f_{1}(F_{\phi}))_{3~3}~=~1- {F_{k}\over {2~2!}} + {3F_{k}^{2}
\over {2~4!}} - {3~F_{k}^{3}\over {6!}}.............\nonu \\
- {4~F_{\eta}\over {3~2!}} + {16~F_{\eta}^{2}\over {3~4!}} -
{320~F_{\eta}^{3}\over {9~6!}}..........\nonu \\
+F_{\eta}({2~F_{\pi}\over {4!}} - {25~F_{\pi}^{2}\over
{3~6!}})..........\nonu \\
-F_{\eta}^{2}~{5~F_{\pi}\over {6!}}................
+F_{k}~({F_{\pi}\over {4!}} - {5~F_{pi}^{2}\over
{3~6!}})........ \nonu \\
-F_{k}^{2}~{7~F_{\pi}\over {2~6!}}..............
+F_{\eta}({3~F_{k}\over {4!}} - {15~F_{k}^{2}\over
{2~6!}})......... \nonu \\
-19~F_{\eta}^{2}{F_{k}\over {6!}} -
{2~F_{\pi}F_{k}F_{\eta}\over {6!}}................ \nonu \\
+O(F_{\phi}^{4})
\end{eqnarray}
As stated in the text, we find that $F_{k} = F_{\eta} = 0$ for
the range of density ($\sim 15- 20$ times normal nuclear matter
density ) and temperature ($\sim 3- 4$ times the critical
temperature) considered. This implies that in the above
expression, the terms with only $F_{\pi}$ will contribute. The
resultant terms can be summed up to give the new masses of $u$,
$d$ and $s$ quarks. So, for flavour asymmetric quark matter the
nonzero corrections to quark mass ($u$ and $d$) terms are
\begin{eqnarray}
(f_{1}(F_{\pi}))_{11}~=~(f_{1}(F_{\pi}))_{2~2}~=~ (1 - F_{\pi})
e^{-F_{\pi}/6}
\end{eqnarray}
Correction to one meson exchange interaction are obtained
from the odd powers in the expansion of $U_{5}$.
\begin{eqnarray}
(f_{I1}(F_{\pi}))_{11}~=~ (1 -
{5~F_{\pi} \over {3~3!}} + {35~F_{\pi}^{2}\over {9~5!}}.....
){\pi^{0}\over f_{\pi}}  \nonu \\
+ ({9~F_{\pi}\over {3~3!}}~+{~75~F_{\pi}^{2}\over {9~5!}}......){\eta\over
{\sqrt{3}}} \nonu \\
(f_{I1}(F_{\pi}))_{22}~=~-~(1 -
{5~F_{\pi} \over {3~3!}} + {35~F_{\pi}^{2}\over {9~5!}}.....
){\pi^{0}\over f_{\pi}} \nonu \\
+ ({9~F_{\pi}\over {3~3!}}~+{~75~F_{\pi}^{2}\over {9~5!}}......){\eta\over
{\sqrt{3}}} \nonu \\
(f_{I1}(F_{\pi}))_{12}~=~(1 -
{5~F_{\pi} \over {3~3!}} + {35~F_{\pi}^{2}\over {9~5!}}.....
){\sqrt{2}~\pi^{+}\over f_{\pi}} \nonu \\
(f_{I1}(F_{\pi}))_{21}~=~(1 -
{5~F_{\pi} \over {3~3!}} + {35~F_{\pi}^{2}\over {9~5!}}.....
){\sqrt{2}~\pi^{-}\over f_{\pi}} \nonu \\
(f_{I1}(F_{\pi}))_{13}~=~(1 -
{F_{\pi} \over {3!}} + {5~F_{\pi}^{2}\over {3~5!}}.....
){\sqrt{2}~k^{+}\over f_{\pi}} \nonu \\
(f_{I1}(F_{\pi}))_{31}~=~(1 -
{F_{\pi} \over {3!}} + {5~F_{\pi}^{2}\over {3~5!}}.....
){\sqrt{2}~k^{-}\over f_{\pi}} \nonu \\
(f_{I1}(F_{\pi}))_{23}~=~(1 -
{F_{\pi} \over {3!}} + {5~F_{\pi}^{2}\over {3~5!}}.....
){\sqrt{2}~k^{0}\over f_{\pi}} \nonu \\
(f_{I1}(F_{\pi}))_{32}~=~(1 -
{F_{\pi} \over {3!}} + {5~F_{\pi}^{2}\over {3~5!}}.....
){\sqrt{2}~\bar k^{0}\over f_{\pi}} \nonu \\
(f_{I1}(F_{\pi}))_{33}~=~0
\end{eqnarray}
The correction to two meson exchange terms come from the even
terms of $U_{5}$.
\begin{eqnarray}
(f_{I2})_{11}=(1-{5~F_{\pi}\over {18}}+ {7~F_{\pi}^{2}\over
{216}})\pi^{+}\pi^{-}+(1-{5~F_{\pi}\over {18}}+ {7~F_{\pi}^{2}\over
{216}}){\pi^{0}}^{2} \nonu \\
+(1- {7~F{\pi}\over {36}}+ {11~F_{\pi}^{2}\over {648}})K^{+}K^{-}
\nonu \\
+(1- {F_{\pi}\over {12}}+ {F_{\pi}^{2}\over {81}})\eta^{2} \nonu \\
(f_{I2})_{22}=(1-{5~F_{\pi}\over {18}}+ {7~F_{\pi}^{2}\over
{216}})\pi^{+}\pi^{-}+(1-{5~F_{\pi}\over {18}}+ {7~F_{\pi}^{2}\over
{216}}){\pi^{0}}^{2} \nonu \\
+(1- {7~F{\pi}\over {36}}+ {11~F_{\pi}^{2}\over {648}})K^{0}\bar {K^{0}}
\nonu \\
+(1- {F_{\pi}\over {12}}+ {F_{\pi}^{2}\over {81}})\eta^{2} \nonu \\
(f_{I2})_{33}=(1- {2~F_{\pi}\over {4!}}+ {10~F_{\pi}^{2}\over
{3~6!}})K^{+}K^{-}
+(1- {2~F_{\pi}\over {4!}}+ {10~F_{\pi}^{2}\over {3~6!}})K^{0}\bar
{K^{0}}
\end{eqnarray}
The correction to meson masses comes through the correction to
kinetic energy term.
\begin{eqnarray}
f_{2\pi}~=~1- {F_{\pi}\over {9}}+ {4~F_{\pi}^{2}\over
{27~3!}}...... \nonu \\
f_{2k}~=~1- {7~F_{\pi}\over {12}}+ {11~F_{\pi}^{2}\over
{108}}...... \nonu \\
f_{2\eta}~=~0
\end{eqnarray}
The $f_{2\eta}~=~0$ implies that there is no change in the
mass of $\eta$ meson in the medium.

\bf {Case b: Three flavour symmetric matter}

Three flavour symmetric matter consist of a single eight
component meson field $\phi$. Also all the quark masses
become equal. The corrections are obtained by putting $j_{\pi}=
j_{K}= j_{\eta}= 8$.
The correction to mass term is given by,
\begin{eqnarray}
f_{1}(F_{\phi})~=~1- {2F_{\phi}\over {3~2!}} + {5~F_{\phi}^{2}\over
{6~4!}}-{95~f_{\phi}^{3}\over {72~6!}}...............
\end{eqnarray}
The one meson exchange term is given by,
\begin{eqnarray}
f_{I1}(F_{\phi})~=~(1- {10~F_{\phi}\over {8~3!}} +
{380~F_{\phi}^{2}\over {3~8^{2}~5!}}....){\Phi\over {f_{\pi}}}
\end{eqnarray}
The two meson exchange term is
\begin{eqnarray}
f_{I2}= (1- {313~F_{\phi}\over {576}}+ {4859~F_{\phi}\over
{103680}})\phi^{2}
\end{eqnarray}
The correction to mass term is given by,
\begin{eqnarray}
f_{2}(F_{\phi})~=~1- {F_{\phi}\over {8}} + {463\over {1620}}
{F_{\phi}^{2}\over {8^{2}}}....................
\end{eqnarray}
\bf {Case c: Two flavour matter}

In case of two flavour matter, only pions are present
in the medium. So that the meson matrix is now a
$2\times 2$ matrix. The contraction rules are as given for
pions in case (a). The results are same as the one obtaind
for the pions in case (a). All the corrections are now summable
and take the forms as given below.
\begin{eqnarray}
f_{1}~=~(1- F_{\pi})e^{-F/6}  \\
f_{2}~=~{1\over {3}}[2~+~{1.5\over {F_{\pi}}}(1 -e^{-2F/3})]  \\
f_{I1}~=~(1 -{F_{\pi}\over {9}})e^{-F_{\pi}/6}
\end{eqnarray}
where the eqn. (A36), eqn. (A37) and eqn. (A38) are the corrections to
quark mass term, pion mass term and quark- pion interaction term.

\vfil
\eject

\vfill
\eject
\newpage
\begin{figure}
\caption { Baryon density vs. $F_{\pi}$ and quark mass $m_q$ (u and d)
plot {\bf (a)} and {\bf (b)} correspond to $F_\pi$ and $m_q$ for
flavour asymmetric case; {\bf (c)} and {\bf (d)} correspond to
$F_\pi$ and $m_q$ for flavour symmetric case respectively.}
\vskip 0.5in
\caption { Temperature vs. $F_{\pi}$ and quark mass $m_q$ (u and d)
plot {\bf (a)} and {\bf (b)} correspond to $F_\pi$ and $m_q$ for
flavour asymmetric case; {\bf (c)} and {\bf (d)} correspond to
$F_\pi$ and $m_q$ for flavour symmetric case respectively.}
\vskip 0.5in
\caption { Baryon density vs. $m_q$ for different pion mass $m_\pi$,
{\bf (a)} for $m_\pi= 110 MeV$, {\bf (b)} for $m_\pi= 90 MeV$
and {\bf (c)} for $m_\pi= 70 MeV$. }
\vskip 0.5in
\caption { Baryon density vs. Pressure for different
temperatures, {\bf (a)} $T= 0 MeV$, {\bf (b)} $T= 50 MeV$, {\bf
(c)} $T= 80 MeV$ and {\bf (d)} $T= 110 MeV$. }
\vskip 0.5in
\caption { Pressure and energy variation with temperature above
critical temperature for zero chemical potential, {\bf (a)}
$3P/T^4$ and {\bf (b)} $E/T^4$}
\end{figure}
\vfill
\eject
\end{document}